\documentclass[superscriptaddress,prl,aps,twocolumn,showpacs,floats]{revtex4}
\usepackage{graphics}
\begin{document}

\title{Jahn-Teller stabilization of a ``polar'' metal oxide surface: 
Fe$_3$O$_4$(001)}
\author{R. Pentcheva}
 \email{pentcheva@lrz.uni-muenchen.de}
\author{F. Wendler}
\author{H.L. Meyerheim}
\altaffiliation[permanent address: ]{MPI f\"ur Mikrostrukturphysik, 
Weinberg 2, 06120  Halle, Germany}
\author{W. Moritz}
\affiliation{Dept. of Earth and Environmental Sciences, University of 
Munich, Theresienstr. 41, 80333 Munich, Germany}
\author{N. Jedrecy}
\affiliation{Lab. Mineralogie-Cristallographie, Universit\'{e} Paris 6 et 7, 
4 place Jussieu, F-75252 Paris, France}
\author{M. Scheffler}
\affiliation{Fritz-Haber-Institut der Max-Planck-Gesellschaft,
         Faradayweg 4-6, D-14195 Berlin, Germany}
\received{04 June 2004}

\begin{abstract}
Using {\sl ab initio thermodynamics} we compile a phase diagram 
for the surface  of Fe$_3$O$_4$(001) as a function of 
temperature and oxygen pressures. A hitherto ignored polar termination 
with octahedral iron and oxygen 
forming a {\sl wave-like} structure along the $[110]$-direction is 
identified as the lowest energy configuration over 
a broad range of oxygen gas-phase conditions. This novel geometry is confirmed 
in a x-ray diffraction analysis. The stabilization of the 
Fe$_3$O$_4$(001)-surface goes together with dramatic changes 
in the electronic and magnetic properties, {\sl e.g.}, a halfmetal-to-metal 
transition. 
\end{abstract}
                

\maketitle
 The surface composition and structure of a multicomponent material is 
a fundamental property 
that has significant consequences for its surface reactivity,  
mechanical, and magnetic properties.  Magnetite is not only important in 
geophysics and mineralogy, but is also attracting increasing attention 
as a potential material for spintronic devices~\cite{kelly,hibma},
due to its half-metallic behavior, coupled with a high Curie
temperature of 858~K.

\begin{figure}[b!]
\scalebox{0.35}{\includegraphics{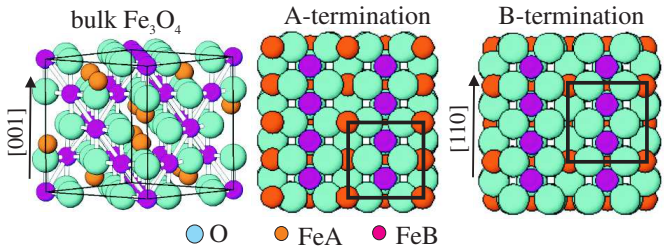}}
\caption{\label{fig-str1x1} {\small The inverse spinel structure of 
magnetite together with a top view of the two bulk truncations of 
Fe$_3$O$_4$(001) with an A- and B-layer, respectively. 
Oxygen atoms, Fe$_{\rm B}$ and Fe$_{\rm A}$ are marked by 
white, grey, and black circles.
} }
\end{figure}
Magnetite crystallizes in the inverse spinel structure, where the oxygen atoms 
form a slightly distorted face centered cubic lattice and the 
iron atoms occupy tetrahedral and octahedral interstitial sites.
 The stacking sequence in the $[001]$-direction consists of A-layers 
containing tetrahedral iron (Fe$_{\rm A}$) and B-layers with oxygen 
and octahedral iron (Fe$_{\rm B}$) atoms ({\sl cf.} Fig.~\ref{fig-str1x1}).   
 
Two alternative approaches are widely used to 
assess possible terminations for metal oxide surfaces, an ionic scheme 
considering the electrostatic energy~\cite{tasker} and the autocompensation 
rule applied to covalent bonds~\cite{lafemina}.  
After the scheme of Tasker~\cite{tasker}, bulk truncations of 
the Fe$_3$O$_4$(001)-surface should
be unstable because of a diverging surface energy due to uncompensated dipole 
moments. Such structures were, therefore, so far discarded. 
Both the ionic~\cite{tasker} and the covalent~\cite{lafemina} approach 
rely on the information 
from the bulk only and suggest that stabilization of a polar surface can 
only be achieved through massive changes of the surface stoichiometry.

For Fe$_3$O$_4$(001) a $(\sqrt{2}\times \sqrt{2})R45^{\circ}$-reconstruction 
has been observed experimentally~\cite{Tar93,Gain97,Cha00,Mij01,Sta00,voogt}.
However, the surface structure and stoichiometry are still under debate. 
Applying the autocompensation rule~\cite{lafemina} to Fe$_3$O$_4$(001), 
 two models for the  
 $(\sqrt{2}\times \sqrt{2})R45^{\circ}$-reconstruction have been proposed 
so far: an A-layer where half of the tetrahedral iron is 
missing~\cite{Tar93,Cha00,Mij01} and a 
B-layer with oxygen vacancies or hydroxyl groups~\cite{voogt,Sta00}. 
Several variations of the former model have been suggested: 
The x-ray photoelectron diffraction (XPD) analysis of Chambers   
{\sl et al.}~\cite{Cha00} and the low energy ion scattering (LEIS) results 
of Mijiritskii {\sl et al.}~\cite{Mij01} were interpreted as a half-filled 
A-layer with strong inward relaxations of the surface layers. Concerning the 
magnitude of the interlayer distances, there is a considerable discrepancy. 
In a molecular dynamics calculation with classical potentials  - the only 
theoretical work so far - Rustad 
{\sl et al.}\cite{Rustad} suggested a 0.5ML A-termination where the 
surface and half of the subsurface Fe$_{\rm A}$ relax in octahedral 
positions in the surface B-layer. 
Also non-autocompensated terminations with a charge ordered B-layer were 
reported from scanning tunneling microscopy (STM)-measurements using 
a ferro-~\cite{wiesendanger92} or an 
antiferromagnetic tip~\cite{mariotto}. 
However, the observed magnetic contrast alone cannot explain 
the LEED- and XRD-superstructure intensities and does not provide a clue on 
the stabilization mechanism of the surface.

In this Letter we present the results of a systematic investigation of 
the composition, structure and properties of the Fe$_3$O$_4$(001)-surface 
based on  density-functional theory (DFT) calculations. Using  
{\sl ab initio atomistic thermodynamics}~\cite{Weinert,Reuter}, we predict 
that a ``polar'' termination is the lowest energy configuration over 
the entire range of accessible oxygen pressures.  The 
stabilization of the surface involves a fundamentally different mechanism, 
which has not been considered so far: While 
most of the previous studies proposed an ordering of surface vacancies
as the origin of the surface reconstruction, here it is 
explained as a Jahn-Teller distortion of the surface atoms forming a 
{\sl wave-like} pattern along the $[110]$-direction. A  
x-ray diffraction (XRD)-analysis supports this theoretically predicted model.

We use the 
full-potential augmented plane waves (FP-APW) method in the 
\textsf{WIEN2k}-implementation~\cite{wien} and the generalized gradient 
approximation (GGA)~\cite{pbe96}.
The Fe$_3$O$_4$(001)-surface is modeled by a 
symmetric slab containing five B- and four to 
six A-layers depending on the structural model. The vacuum 
between the repeated slabs amounts to $10$~\AA.  
We have ensured convergence 
with respect to the thickness of the slab by repeating the calculations 
 for the most stable configurations with a thicker slab with an 
additional A- and B-layer on both sides (approximately 100 atoms in 
the unit cell as opposed to 70 in the original unit cell). 
The surface free energies for both supercells 
are within $\pm 3$meV/\AA$^2$ equal.  The lateral parameter of the supercell 
is set to the GGA bulk lattice constant, $8.42$~{\AA},  which is in good 
agreement with the experimental value  of $8.394$~{\AA}. 

The convergence parameters for the mixed APW+lo and LAPW basis set are: 
$R^{\rm MT}_{\rm Fe}=1.90\,\textrm{bohr}$, 
$R^{\rm MT}_{\rm O}=1.40\,\textrm{bohr}$, inside the muffin tins (MTs) 
wave functions expansion 
in spherical harmonics up to $l_{\rm max}^{\rm wf}=10$ and non-spherical
contributions to the electron density and potential up to 
$l_{\rm max}^{\rm pot.}=4$. The energy cutoff for the plane wave 
representation in the interstitial is $E_{\rm max}^{\rm wf}=19$~Ry 
for the wave functions and $E_{\rm max}^{\rm pot.}=196$~Ry for the 
potential. With these cutoff parameters a convergence of energy differences 
better than $1$~mRy  is achieved.
Results for the $(\sqrt{2}\times \sqrt{2})R45^{\circ}$-unit cell 
are obtained with 4 $k_{\parallel}$-points in the irreducible 
part of the Brillouin zone.

\begin{figure}[t!]
\scalebox{0.35}{\includegraphics{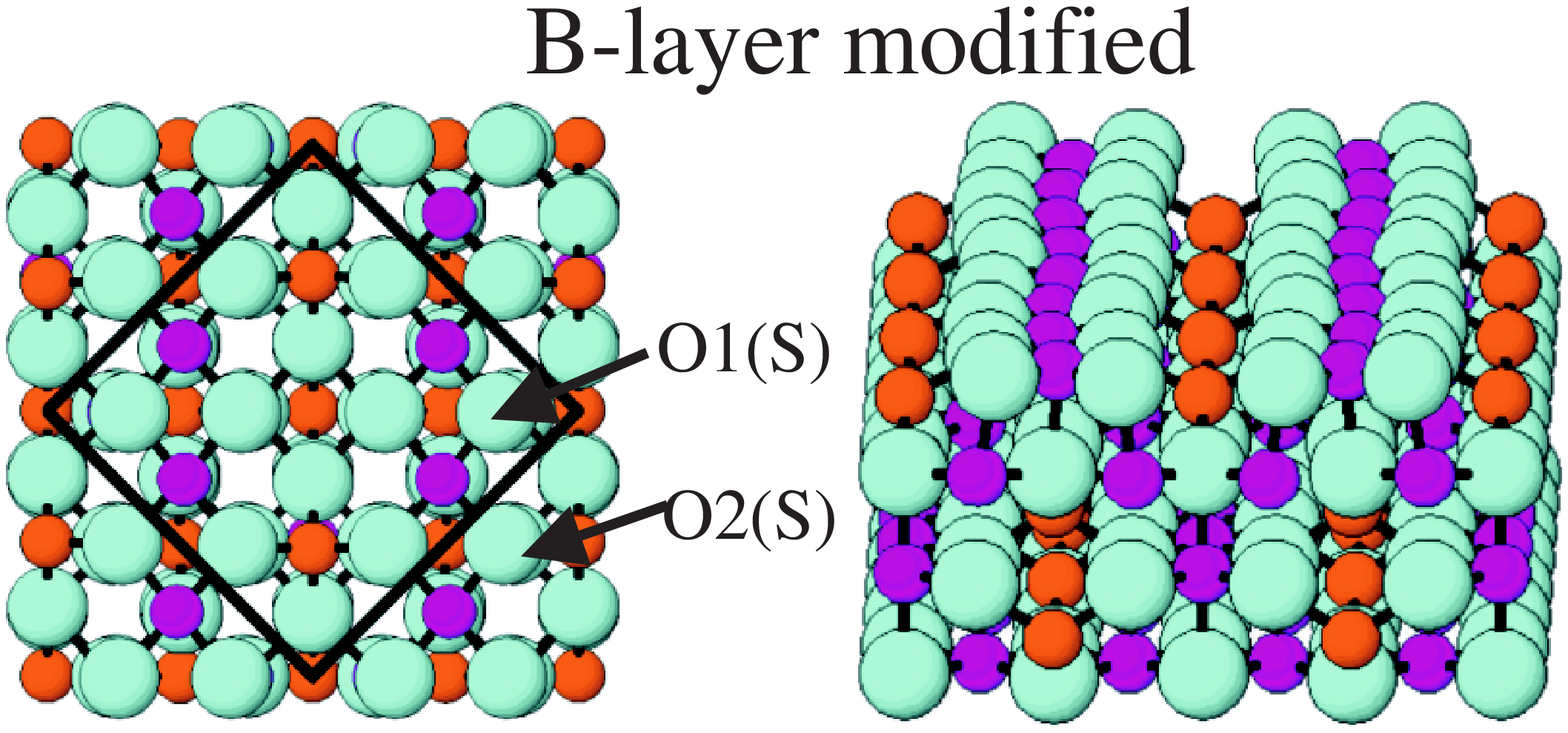}}
\scalebox{0.78}{\includegraphics{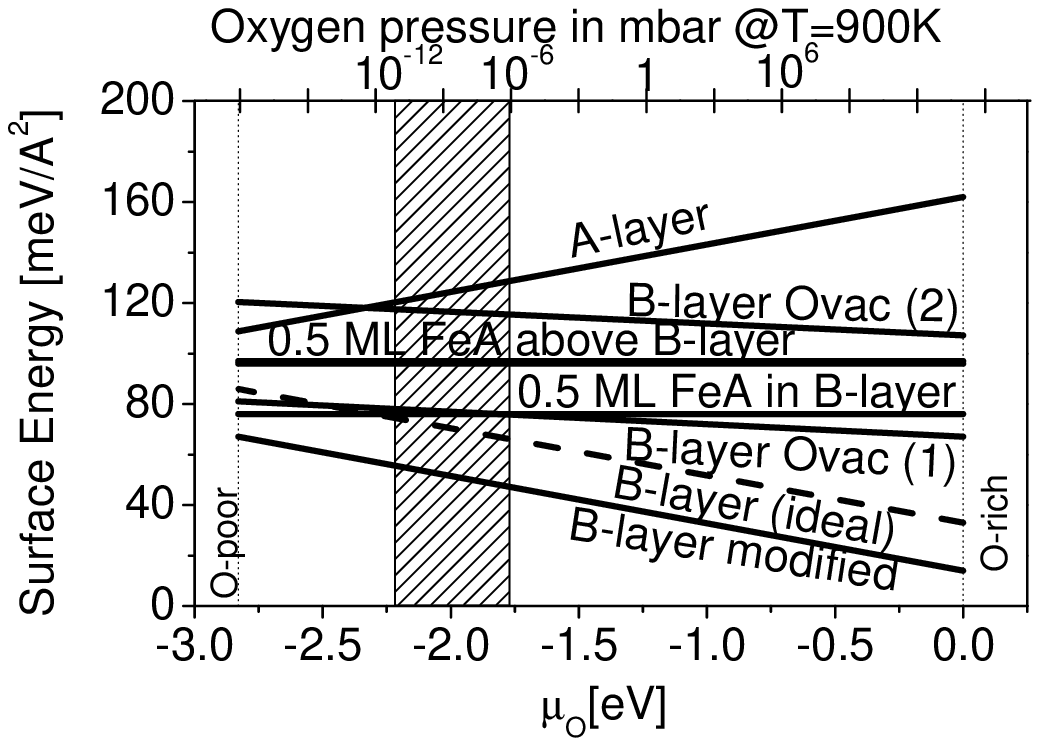}}
\caption{\label{fig-enmpt} {\small Calculated surface free energy $\gamma(T,p)$
               as a function of the chemical potential of oxygen
               (bottom $x$-axis)  for all studied terminations.  
               In the top $x$-axis the $\mu_O$ is converted into a 
               pressure scale at $T=900$~K. The vertical 
               lines mark the 
               {\sl oxygen-poor} and {\sl oxygen-rich} limits of the oxygen 
               chemical potential, with $1/2 E_{O_2}^{total}$ used as
               zero reference. The dashed 
               region marks the range of pressures that were used during 
               sample preparation in the experiment. The 
               B-termination with bulk atomic positions and  
               modified positions with
               a $(\sqrt{2}\times \sqrt{2})R45^{\circ}$-periodicity are 
               marked as ideal and modified, respectively. The 
               B-layer with oxygen vacancies above an octahedral iron and 
               next to a tetrahedral iron are denoted by (1) and (2), 
               respectively. Top and side view of the modified B-layer is  
               given at the top. Surface oxygens with and without a 
               Fe$_{\rm A}$ neighbor are denoted by O2(S) and O1(S). 
               For the color code see Fig.~\ref{fig-str1x1}. 
} }
\end{figure}

The configurations we have considered include  
{\sl (i)} a 0.5 ML of tetrahedral iron above the 
B-layer~\cite{Tar93,Cha00, Mij01}, 
{\sl (ii)} or relaxed in the B-layer; 
{\sl (iii)} a  
      modified 0.5 ML A-termination where the surface and half of the 
      subsurface Fe$_{\rm A}$ occupies 
      octahedral sites in the B-layer~\cite{Rustad}. Additionally, we have 
considered two geometries with one oxygen vacancy per unit cell: either 
as a nearest neighbor 
to a tetrahedral iron or above an octahedral iron. 
Together with the non-autocompensated A- and 
B-terminations shown in Fig.{\ref{fig-str1x1} they were used 
as a starting point for the DFT-calculations, where we performed a full 
structural optimization of the atomic positions in the outer two 
AB-double-layers with damped Newton dynamics~\cite{Koh96}.

The calculated surface phase diagram for all studied  terminations is compiled 
in Fig.~\ref{fig-enmpt}. The oxygen chemical potential is converted 
into oxygen pressures 
for a fixed temperature of $T=900$~K, which corresponds 
to a typical annealing temperature in experiment. 
The range of allowed oxygen chemical potentials lies between the 
``oxygen-poor''-limit,  which corresponds to the case where the oxide would 
decompose in iron crystallites and oxygen gas and the ``oxygen-rich''-limit, 
defined by the chemical potential of oxygen in the oxygen molecule, set as 
zero reference.  In Fig.~\ref{fig-enmpt} the surface energy is given 
with respect to bulk magnetite. We note that 
at $T=900$~K and $p_{O_2}>10^{-6}$~mbar hematite becomes the 
thermodynamically stable phase~\cite{Richard}. Taking into account the 
calculated enthalpy 
of formation for magnetite (11.32~eV) and hematite (8.024~eV)~\cite{Wang98}, 
the stability range of  Fe$_3$O$_4$ extends up to $\mu_{O}=-1.5$~eV. 

Autocompensated terminations, {\sl e.g.}, the ones containing 
0.5 ML Fe$_{\rm A}$ in the surface layer, are 
independent of the oxygen pressure, while the effect of the oxygen 
environment is most distinct 
for the two bulk truncations with an A- or a B-layer. In line with  
electrostatic considerations, the A-termination is unstable and this 
trend is substantially enhanced for {\sl oxygen-rich} conditions.  
A most striking result is that the other bulk truncated surface, the 
B-termination, which has been so far discarded as a non-compensated surface,
turns out to be the most stable configuration over a 
broad range of oxygen pressures even in the ideal (unrelaxed) geometry. 
For {\sl oxygen-poor} conditions 
it competes with a 0.5 ML tetrahedral Fe 
relaxed in the B-layer and with a B-termination with O-vacancies lying above 
an octahedral iron in the subsurface B-layer. 
\begin{figure}[t!]
\scalebox{1.0}{\includegraphics{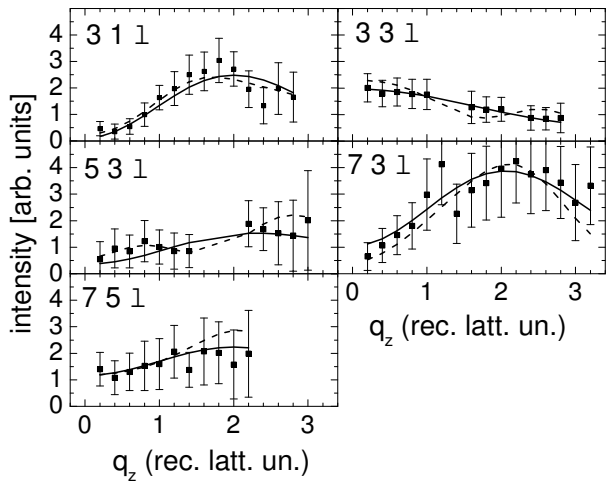}}
\caption{\label{fig-xrd} {\small Results of the XRD-analysis. The 
    experimental data is marked by squares. Solid and dashed lines 
correspond to fits assuming the modified B-termination  and 0.5 ML 
Fe$_{\rm A}$-termination, respectively. 
    } }
\end{figure}
An asymmetric force on the surface oxygens in the unreconstructed ideal 
B-termination entails a {\sl wave-like}-distortion along the 
Fe$_{\rm B}$-row. Hence, the origin of the symmetry 
reduction and the observed $(\sqrt{2}\times \sqrt{2})R45^{\circ}$-pattern 
is a Jahn-Teller-distortion~\cite{jt37,cusp}. Compared to the ideal 
B-termination, the formation of
 this {\sl modified} B-termination is connected with an energy 
gain of approximately $20$~meV/\AA$^2$ and a reduction of the work function 
of $0.46$~eV from $5.78$~eV to $5.32$~eV. We note that a reduction 
of  $\Phi$ indicates a decrease of the dipole moment induced on the surface, 
{\sl i.e.}, a decrease in polarity.  The calculated work function is 
in good agreement with the value measured in photoemission 
experiments~\cite{guenth03}.

A  top and side view of the novel surface geometry, is  given at the top of
 Fig.~\ref{fig-enmpt}. The {\sl wave-like}-pattern along the 
$[110]$-direction emerges from a strong 
lateral relaxation of $0.30$~\AA\ of one of the two oxygen atoms with a
 missing Fe$_{\rm A}$-neighbor (marked as O1(S) in Fig.~\ref{fig-enmpt}) 
towards the Fe$_{\rm B}$-row and a subsequent alternating shift 
of Fe$_{\rm B}$-pairs by $\pm 0.09$~\AA\ perpendicular to the 
Fe$_{\rm B}$-row. The oxygen atoms with a Fe$_{\rm A}$-neighbor (O2(S)) 
relax by $0.14$~\AA\ towards and by $0.03$~\AA\ parallel to the 
Fe$_{\rm B}$-row. Due to the lateral relaxation of oxygen the 
Fe$_{\rm B}$-oxygen distance is reduced from 2.06~\AA\ in the 
bulk to 1.96~\AA\ on the surface and are thus closer to the bulk values
 for the tetrahedral ($d_{\rm O-Fe_A}^{\rm bulk}=1.89$~\AA) coordination.
 Perpendicular to the surface, all surface atoms 
relax inwards except for O1(S), which relaxes outwards. The atoms 
in the subsurface A- and B-layer relax outwards. A detailed structural 
analysis will be presented elsewhere~\cite{rp04}.

X-ray measurements were performed on a natural crystal~\cite{purity} 
at the beam line DW12 at 
LURE/Orsay with a wave length of 0.826~\AA. In total, 220 symmetrically 
independent reflections were measured at room temperature
in 5 crystal truncation rods (CTR) and 6 superstructure rods. 
The cleanliness of the sample and the reproducibility of 
the $(\sqrt{2}\times \sqrt{2})R45^{\circ}$-LEED pattern were verified 
by a repeated sputtering and annealing up to 
850 K in UHV and oxygen pressures up to $5\times10^{-6}$~mbar. 
Furthermore, the LEED I/V curves of the natural 
and an artificial crystal are nearly identical~\cite{frank}.   

To minimize the effect of surface inhomogeneities only the superstructure 
reflections were considered in the analysis. The experimental data and the 
calculated curves for the two most promising terminations, {\sl i.e.}, the 
half-filled A-layer of  tetrahedral iron and the modified 
B-termination are displayed in Fig.~\ref{fig-xrd}. 
The corresponding R-factors 
obtained after a structural refinement with the DFT-geometries are 
$0.24$ and $0.19$, respectively, which indicates a 
preference for the modified B-termination. Both 
models vary only by the additional tetrahedral iron in the half-filled 
A-layer. The relaxations of the remaining atoms are rather similar leading to 
an overall similarity in the intensity distribution along $q_z$. Still the 
fit with the modified B-layer is noticeably better. 
The B-termination with oxygen vacancies, 
proposed by Stanka {\sl et al.}~\cite{Sta00} is incompatible with 
the XRD-data (R-factor worse than 0.6). The {\sl wave-like} 
structure along the $[110]$-direction observed in the  
STM-measurements of Stanka {\sl et al.}~\cite{Sta00} is a further 
evidence for the modified B-termination~\cite{STM}.

The stabilization of the Fe$_3$O$_4$(001)-surface and the unusual 
lattice distortions are connected with strong 
 changes in the electronic and magnetic properties. 
In the bulk, magnetite shows 
a {\sl half-metallic} behavior with a band gap in the majority spin 
channel of approximately 0.5~eV and a $100\%$ spin-polarization due 
to the $t_{2g}$-states of Fe$_{\rm B}$ at $E_F$ in the minority spin 
channel~\cite{satpathy}. 

\begin{figure}[t!]
\scalebox{0.72}{\includegraphics{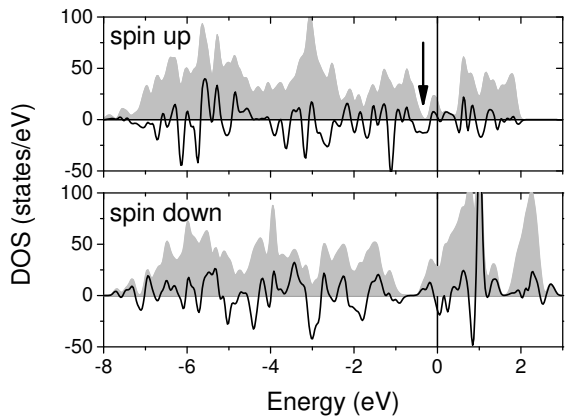}}
\caption{\label{fig-dosbulk} {\small Total density of states (DOS) 
                of Fe$_3$O$_4$(001) terminated by a modified B-layer 
                for the majority (minority) spin channel in the upper 
                (lower) panel is shown in grey. 
                The difference of the DOS of the modified B-termination 
                and the bulk-terminated B-layer is given with a black line. 
                A Gauss broadening of $2\sigma=0.05$~eV is used. 
} }
\end{figure}
On the surface (cf. Fig.~\ref{fig-dosbulk}) surface states in 
the majority spin gap reduce the spin-polarization at the Fermi level 
to approximately $60\%$. Spin-resolved photoemission 
measurements support this result~\cite{guenth03}. Since electrons near the 
Fermi level are important for transport properties, we examined closer
their nature: the states at $E_F$ are confined to  the surface 
layer and are due to the hybridization between $d_{x_2-y_2}$-states of
Fe$_{\rm B}$ and $p_x,p_y$-states of O1(S). Additionally, there is  
a noticeable contribution from the tetrahedral iron in the subsurface layer.
In order to get more insight into the driving mechanisms towards the 
reconstruction of the surface we have plotted in Fig.~\ref{fig-dosbulk} 
also the difference between the DOS of the B-termination with modified and 
ideal positions of the atoms. For the modified 
B-termination a band gap opens in the majority spin channel 
between $-0.4$ and $-0.2$~eV where states of $d_{x_2-y_2}$ ($d_{z_2}$)
character are shifted 
to higher (lower) energies. The latter cross the Fermi level 
and eventually become partially unoccupied.  Vice versa, in the minority 
spin channel, bands become occupied. Altogether this reduces the energy 
of the system. 

This combination of a Jahn-Teller-distortion and a spin-flip 
leads to interesting magnetic 
properties. In the bulk magnetite is a ferrimagnet with the magnetic moments 
of tetrahedral ($-3.43\,\mu_{\rm B}$) and octahedral iron ($3.50\,\mu_{\rm B}$)
oriented antiparallel  to each other and 
a total magnetic moment of $4.0\,\mu_{\rm B}$ per formula unit. In the modified
B-termination the magnetic moments of the subsurface Fe$_{\rm A}$ ($-3.45\,\mu_{\rm B}$) is close to the bulk value. However, the magnetic moment of the 
surface Fe$_{\rm B}$ ($3.00\,\mu_{\rm B}$) is substantially 
reduced due to the strong relaxations.

Additionally, a 
substantial magnetic moment of approximately $0.25\,\mu_{\rm B}$ 
is induced in the undercoordinated surface oxygen  with a missing bond to 
tetrahedral iron, while the magnetic moments of the rest of the surface oxygen 
atoms are close to the bulk value, $0.07\,\mu_{\rm B}$. A similar magnetization was 
observed for the oxygen-terminated  
$\alpha$-${\rm Fe}_2{\rm O}_3(0001)$-surface~\cite{Wang98}.  

Based on {\sl ab initio thermodynamics} we identify a B-termination with 
octahedral iron and oxygen in the surface layer forming a {\sl wave-like} 
structure along the [110]-direction as the thermodynamically stable 
configuration over a broad range of oxygen pressures. Instead of ordering 
of surface defects the symmetry reduction  is achieved by a Jahn-Teller 
distortion  
 and is connected with strong changes 
of the electronic and magnetic properties, {\sl e.g.} a metalization 
at the surface.   
Experimental evidence for the DFT-predicted 
geometry is given in STM-measurements~\cite{Sta00,guenth03} and a 
XRD-analysis performed with the atomic positions obtained with DFT.

Theoretically, ``polar'' surfaces have been 
predicted for a number of metal oxide surfaces, e.g. 
Fe$_2$O$_3(0001)$~\cite{Wang98}, RuO$_2$(110)~\cite{Reuter}, 
ZnO(0001)~\cite{Kresse03}, and PdO~\cite{Rogal}. The modified B-termination 
of Fe$_3$O$_4$(001) is a further example that excluding such surfaces 
from the structural analysis can be misleading.

This work is supported 
by the DFG, PE 883.


\begin{thebibliography}{99}
\bibitem{cusp}     The total energy as a function of 
                   the distortion shows a cusp at the symmetric initial 
                   configuration and is thus in accord with the 
                   Jahn-Teller theorem.
\bibitem{purity}   Besides $0.02\%$ Mn no contaminations of the crystal 
                   were detected with x-ray fluorescence spectroscopy.
\end{thebibliography}
\end{document}